\documentclass[%
 reprint,
 amsmath,amssymb,
 aps,floatfix,superscriptaddress,
nofootinbib]
{revtex4-2}

\usepackage{graphicx}
\usepackage{dcolumn}
\usepackage{bm}
\usepackage{epstopdf}
 \usepackage[usenames,dvipsnames]{pstricks}
 \usepackage{epsfig}
 \usepackage{pst-grad} 
 \usepackage{pst-plot} 
 \usepackage[space]{grffile} 
 \usepackage{etoolbox} 
 \makeatletter 
 \patchcmd\Gread@eps{\@inputcheck#1 }{\@inputcheck"#1"\relax}{}{}
 \makeatother
\usepackage{booktabs}
\usepackage{boxhandler}
\usepackage{amsmath}
\usepackage{subcaption}
\usepackage{braket}
\usepackage[cmintegrals]{newtxmath}
\usepackage{multirow}
\usepackage{enumitem}

\newcommand{\overbar}[1]{\mkern 1.5mu\overline{\mkern-1.5mu#1\mkern-1.5mu}\mkern 1.5mu}

\newcommand{\centra}{$\mathcal{\overline{Z}(\overline{S})}$ }

\begin{document}

\title{On the logical error rate of sparse quantum codes}

\author{Patricio Fuentes $^{1,\ *}$, {Josu Etxezarreta Martinez}$^{1,\ *}$, Pedro M. Crespo} 
\email[Correspondence: ]{\{pfuentesu, jetxezarreta, pcrespo\}@tecnun.es}
    \affiliation{ Tecnun - University of Navarra, 20018 San Sebastian, Spain}

\author{Javier Garcia-Frias}
        \email[Correspondence: ]{jgf@udel.edu}
    \affiliation{University of Delaware, Newark, 19716 Delaware, USA}

\date{\today}

\begin{abstract}
The quantum paradigm presents a phenomenon known as degeneracy that can potentially improve the performance of quantum error correcting codes. However, the effects of this mechanism are sometimes ignored when evaluating the performance of sparse quantum codes and the logical error rate is not always correctly reported. In this paper, we discuss previously existing methods to compute the logical error rate and we present an efficient coset-based method inspired by classical coding strategies to estimate degenerate errors and distinguish them from logical errors. Additionally, we show that the proposed method presents a computational advantage for the family of Calderbank-Shor-Steane (CSS) codes. We use this method to prove that degenerate errors are frequent in a specific family of sparse quantum codes, which stresses the importance of accurately reporting their performance. Our results also reveal that the modified decoding strategies proposed in the literature are an important tool to improve the performance of sparse quantum codes.

\end{abstract}

\keywords{Quantum error correction, quantum low density parity check codes, quantum low density generator matrix codes, iterative decoding.}
\maketitle

\section{Introduction}
\label{sec:introduction}
When quantum stabilizer codes built from sparse classical codes are employed in the quantum paradigm, they are impacted by a phenomenon known as degeneracy \cite{degen1, degen2, degen3, degen4}, which has no classical equivalent. This causes stabilizer codes to exhibit a particular coset structure in which multiple different error patterns act identically on the transmitted information \cite{symplec1, symplec2, CSS2}. 
The manifestation of degeneracy in the design of sparse quantum codes and its effects on the decoding process has been studied extensively \cite{degen3, degen4, QSC, logical, softVit, neural, Hard, softPoul, reviewPat}. Unfortunately, although degeneracy may potentially improve performance, limited research exists on how to quantify the true impact that this phenomenon has on Quantum Low Density Parity Check (QLDPC) codes. This has resulted in the performance of QLDPC codes being assessed differently throughout the literature; while some research considers the effects of degeneracy by computing the metric known as the \textit{logical error rate} \cite{degen3, refined, osd1, osd2, exploiting, Til, mod-BP}, other works employ the classical strategy of computing the \textit{physical error rate} \cite{jgf1, jgf2, qldpc15, efb, bab1, patrick1}, a metric which provides an upper bound on the logical error rate of these codes since it ignores degeneracy. In the context of degenerate quantum codes, the discrepancy between results computed based on the physical error rate and the logical error rate can become significant. QLDPC codes (also commonly referred to as sparse quantum codes) are famous for their degenerate nature \cite{degen3, degen4, Hard, reviewPat, QEClidar, trapping-sets}, which makes it important to accurately asses their performance using the logical error rate.

The logical error rate is a well-known metric and has been widely employed to assess the performance of other families of Quantum Error Correction (QEC) codes such as Quantum Turbo Codes (QTC) and quantum topological codes \cite{softVit, EAQTC, toricphd1, toricphd2, sabo}. For these particular error correction schemes, the decoders that are employed are sometimes capable of distinguishing between error equivalence classes (they can account for some aspects of degeneracy), which makes it easy to compute the logical error rate. In contrast, the classical decoding algorithms \cite{BP, spa} that are used to decode sparse quantum codes are unable to account for the presence of degenerate error operators, and so their logical error rate must be computed differently. 

  Given the coset structure of sparse quantum codes, intuition would point towards approaching the issue of calculating the logical error rate by finding and comparing the stabilizer cosets of the estimated error sequences and the stabilizer cosets of the channel errors. Unfortunately, the task of computing stabilizer cosets has been shown to be computationally hard \cite{degen4, Hard, efb}, which is the reason why the performance of some sparse quantum codes \cite{jgf1, jgf2, qldpc15, bab1, patrick1, efb} has been assessed based on the physical error rate.  This metric is computed by comparing the error sequence estimated by the decoder, $\hat{\mathbf{E}} \in \mathcal{\overbar{G}}_N$, to the channel error, $\mathbf{E} \in \mathcal{\overbar{G}}_N$, where $\mathcal{\overbar{G}}_N$ denotes the $N$-fold effective Pauli group. Essentially, if the estimation matches the channel error, the decoder has been successful, and if not, a decoding failure has occurred. Note, however, that because this metric ignores the degenerate nature of stabilizer codes, the physical error rate overestimates the number of decoding failures and actually represents an upper bound on the performance of stabilizer codes.

  Despite the use of the physical error rate in some works, other literature has successfully computed the logical error rate of specific sparse quantum codes \cite{degen3, refined, osd1, osd2, exploiting, Til, mod-BP}. These works succeed in computing the logical error rate because they do not approach the issue from the perspective of stabilizer cosets. Instead, in most of these works \cite{degen3, refined, osd1, osd2, exploiting}, Gaussian elimination is used to obtain the parity check matrix of the code in what is known as \textit{standard form} \cite{QSC, Benny, Wilde}, and then this matrix is used to extract a basis for the encoded Pauli operators of the corresponding codes. Then, this basis can be employed to distinguish between degenerate errors and logical errors. Additionally, other research \cite{mod-BP} employs a different, albeit much more computationally demanding, method to compute the logical error rate.

Against this backdrop, in this article we document the methods that have previously been used in the literature to compute the logical error rate of sparse quantum codes. In addition, we introduce our own group theoretic strategy to accurately assess the effects of degeneracy on sparse quantum codes and compute their logical error rate. We also show how for sparse CSS quantum codes the method we propose herein has a computational complexity advantage over those that have been employed previously. Following this, we use our strategy to analyze the frequency with which degenerate errors occur when using a specific family of sparse quantum codes and we provide insight on how the design and decoding of these codes can be improved.

\section{Degeneracy and stabilizer codes} \label{preliminaries}

Throughout this paper we assume that the reader is familiar with basic Quantum Error Correction (QEC) concepts based on physical two-level systems (qubits) such as stabilizer codes, their coset structure, and the different types of possible errors. This section is meant as an introduction to the notation we will use (that of \cite{reviewPat}) and to provide a brief summary of some of the ideas of \cite{reviewPat}. For a complete overview on these preliminaries refer to \cite{reviewPat, QEClidar, NielsenChuang}.

\subsection{Coset structure of the Pauli Group}

The coherence loss suffered by quantum information over time can be approximated as the action of operators that belong to the Pauli Group on the qubits that store this information. In most cases, the action of these operators is modelled by means of an abstraction known as the Pauli channel. For single qubit quantum states, the Pauli channel will act on said states by combining them with an element of the single-qubit Pauli group
$\mathcal{G}_1=(\tilde{\Pi}, \times)$. The set $\tilde{\Pi}$ is given by 

\begin{align*} 
\begin{split} 
        \tilde{\Pi} &= \{\Theta_1 I, \Theta_2 X, \Theta_3 Y, \Theta_4 Z\} \\
        &= \{\pm I, \pm iI, \pm X, \pm iX, \pm Y, \pm iY, \pm Z, \pm iZ\} \ ,
\end{split}
\end{align*}

where $\Theta_l = \{\pm 1, \pm i\}, \{X, Y, Z\}$ are the Pauli matrices, $I$ is the $2\times2$ identity matrix, and $l=1,2,3,4$. The group operation $\times$ represents the well-known product of Pauli matrices, which is written as

\begin{equation}\label{eq:pauli-rel}
\sigma_a\times\sigma_b=I\delta_{a,b}+i\sum_{c=1}^{3}\epsilon_{abc}\sigma_c,
\end{equation}
where $\delta$ represents the Kronecker delta and
\[ \epsilon_{abc} = \left\{ \begin{array}{rl}
         1 & \mbox{if $(a,b,c)=\{(1,2,3),(3,1,2),(2,3,1)\}$},\\
        -1 & \mbox{ if $(a,b,c)=\{(3,2,1),(1,3,2),(2,1,3)\}$},\\
         0 & \mbox{if $a=b$ or $b=c$ or $a=c$.}\end{array} \right. \] 

In \eqref{eq:pauli-rel} we have denoted the $X$ Pauli matrix as $\sigma_1$, the $Y$ Pauli matrix as $\sigma_2$, and the $Z$ Pauli matrix as $\sigma_3$. 

We can generalize the single qubit Pauli group so that it can be applied to $N$-qubit quantum states based on the tensor product. The $N$-qubit Pauli group is by $\mathcal{G}_N = (\tilde{\Pi}^{\otimes N}, \cdot)$. The set $\tilde{\Pi}^{\otimes N}$ is given by 
$$\tilde{\Pi}^{\otimes N} = \{\theta_1I, \theta_2X, \theta_3Y, \theta_4Z\}^{\otimes N},$$ 
where $\theta_k = \{\pm 1, \pm i\}$, and $\otimes$ denotes the tensor product. For all $\mathrm{A},\mathrm{B}\in \mathcal{G}_N$, the group operation $\cdot$ is defined as 

\begin{equation} \label{eq:product-gn}
\mathrm{A}\cdot \mathrm{B}=A_1\times B_1\otimes A_2\times A_2\otimes\ldots \otimes A_N\times B_N .
\end{equation}

Because it makes good physical sense to neglect the global phase of Pauli operators \cite{catalytic}, i.e, disregard the overall factors $\{\pm1, \pm i\}$, it is typical in the field of QEC to work with a reduced version of the Pauli Group known as the \textit{effective Pauli Group}. The $N$-qubit effective Pauli group is denoted by $\overline{\mathcal{G}}_N=(\Pi^{\otimes N}, \star)$, where 

$${\Pi}^{\otimes N} = \{I, X, Y, Z\}^{\otimes N},$$ and $\star$ behaves as in \eqref{eq:product-gn} but with the operation between single Pauli operator products defined not as in \eqref{eq:pauli-rel} but as
\begin{equation} \label{eq:eff-gn-prod}
     \sigma_a\star \sigma_b=I\delta_{a,b}+\sum_{c=1}^{3}|\epsilon_{abc}|\sigma_c,
\end{equation}
where $\epsilon_{abc}$ is the same as in \eqref{eq:pauli-rel}.

Thus, against this backdrop, efficient QEC strategies for general Pauli channels are defined as those methods that are capable of reverting the action of elements that belong to the effective Pauli group. It is important to note, however, that $\overline{\mathcal{G}}_N$ is abelian whereas $\mathcal{G}_N$ is not, which means that the commutation relations that exist between Pauli operators (see equation \eqref{eq:product-gn}) are lost between effective Pauli operators (see equation \eqref{eq:eff-gn-prod}). Given the importance of these commutation relations for the purposes of error correction \cite{reviewPat,NielsenChuang,EAQECC}, it is important that they be recovered. For this purpose we define the symplectic map $\beta:\Pi^{\otimes N} \rightarrow \mathbb{F}_2^{2N}$, which is an isomorphism between the group $\overline{\mathcal{G}}_N=(\Pi^{\otimes N},\star$) and the group ($\mathbb{F}_2^{2N}, \oplus$) of $2N$ binary-tuples under the mod 2 sum operation. For clarity, throughout the remainder of this paper, lower case boldface romans without a subscript will be used to denote $2N$ binary-tuples that belong to $\mathbb{F}_2^{2N}$, and lower case boldface romans with a subscript will be used to denote $N$ binary-tuples that belong to $\mathbb{F}_2^{N}$. The symplectic mapping $\beta$ is defined $\forall \mathbf{A} \in \overline{\mathcal{G}}_N$ as

\[\beta(\mathbf{A})=\mathbf{a}=(\mathbf{a}_x|\mathbf{a}_z),\;\mathbf{a}_x,\mathbf{a}_z\in \mathbb{F}_2^N,\]
where the values of the entries of $\mathbf{a}_x$ and $\mathbf{a}_z$ at position $i=1,\ldots, N$, are directly dependent on the single qubit Pauli operator, $[\mathbf{A}]_i$, located at the $i$-th position in the tensor product that makes up $\mathbf{A} \in \overline{\mathcal{G}}_N$.

Note that the commutation properties of the Pauli operators with regard to the $\cdot$ product are not recovered by just defining the isomorphism $\beta$ (after all, ($\mathbb{F}_2^{2N}, \oplus)$ is also an abelian group). Thus, we define the symplectic product, $\mathbf{a}\odot \mathbf{b}\in \mathbb{F}_2 \ \forall \ \mathbf{a},\mathbf{b}\in  \mathbb{F}_2^{2N}$ as 
\begin{equation}\label{eq:car34}
\mathbf{a}\odot \mathbf{b}\triangleq (\mathbf{a}_x\circledast \mathbf{b}_z)\oplus(\mathbf{a}_z\circledast \mathbf{b}_x).
\end{equation}
where $\circledast$ is the standard mod 2 inner product defined on  $(\mathbb{F}_2^{N}, \oplus)$ considered as a vector space over the field $\mathbb{F}_2$. The symplectic product tells us if any two operators $\mathbf{A},\mathbf{B}$ in $\overline{\mathcal{G}}_N$ will either commute or anticommute with regard to the group operation $\cdot$ in ${\mathcal{G}}_N$, if and only if
the symplectic product between $\beta(\mathbf{A})=\mathbf{a}$ and $\beta(\mathbf{B})=\mathbf{b}$, i.e., $\mathbf{a}\odot \mathbf{b}$, takes the value 0 or 1, respectively \normalfont{\cite{josurev, catalytic}}.

\subsection{Stabilizer Codes}

Stabilizer codes are a widely extended family of QEC codes. They are constructed based on a set of generators that define the so-called stabilizer group $\mathcal{\overline{S}}\subset\mathcal{\overline{G}}_N$. The stabilizer generators are a set of operators $\{\mathbf{S}_v\}_{v=1}^{N-k}$ that belong to $\overline{\mathcal{G}}_N$ and whose sympletic representatives commute with regard to the symplectic product. This means that the stabilizer is comprised of elements of the effective Pauli group that commute amongst themselves with regard to $\cdot$ in $\mathcal{G}_N$. The stabilizer $\mathcal{\overline{S}}$ has $2^{N-k}$ distinct elements and is completely represented by its $N-k$ independent generators (the rest of the stabilizer elements are combinations of the elements of the minimal set). The codespace defined by the stabilizer, and hence a stabilizer code, can be described mathematically as

\begin{equation}
\label{eq:stabilizer}
    \mathcal{C}(\mathcal{\overline{S}}) = \{|\psi\rangle \in \mathcal{H}_2^{\otimes N}:\mathbf{S}_i|\psi\rangle=|\psi\rangle,i=1 \dots N-k\},
\end{equation}
where $\mathbf{S}_i|\psi\rangle\in \mathcal{H}_2^{\otimes N}$ is the evolution of state $|\psi\rangle$ under stabilizer generator $\mathbf{S}_i$. Note that $\mathcal{C}(\mathcal{\overline{S}})$ is the subspace of $\mathcal{H}_2^{\otimes N}$ formed by the simultaneous $+1$-eigenspaces of all the operators in the stabilizer group $\mathcal{\overline{S}}$. Here $\mathcal{H}_2^{\otimes N}$ denotes the complex Hilbert space of dimension $2^{N}$ that comprises the state space of $N$-qubit systems.

It is also possible for elements of $\mathcal{\overline{G}}_N$ to commute with regard to the symplectic product with the elements of $\mathcal{\overline{S}}$ but not to belong to $\mathcal{\overline{S}}$. These elements are important because they act non-trivially on encoded quantum states (they corrupt the encoded quantum information but map the codespace to itself). Together with the elements of $\mathcal{\overline{S}}$, these operators define the group ${\overline{\mathcal{Z}}(\overline{\mathcal{S}})} \subset \mathcal{\overline{G}}_N$ known as the effective centralizer. This group is the set of all operators in $\overline{\mathcal{G}}_N$ whose symplectic representatives commute with regard to the symplectic product with all the symplectic representatives of the stabilizer generators $\beta({\mathbf{S}_v})\in \mathbb{F}_2^{2N}$.  

Based on the relationship between the stabilizer and the effective centralizer, the effective centralizer can actually be understood as a union of stabilizer cosets. More specifically, ${\overline{\mathcal{Z}}(\overline{\mathcal{S}})}$ can be partitioned into $2^{2k}$ cosets of $\mathcal{\overline{S}}$, each of these cosets being indexed by a coset representative $\{\mathbf{L}_j\}_{j=1}^{2^{2k}} \in \mathcal{\overline{Z}(\overline{S})}$. Throughout this paper, we will refer to these representatives as logical operators. Whenever a representative $\mathbf{L}_j$ is multiplied in terms of the group operation over the effective Pauli group, the $\star$ product, by all the elements of $\mathcal{\overline{S}}$, the stabilizer coset, $\mathbf{L}_j\star\mathcal{\overline{S}}$, is obtained. This can be written as 

\begin{equation} \label{centra-cosets}
\overline{\mathcal{Z}}(\overline{\mathcal{S}})=\bigcup_{j=1}^{2^{2k}} \mathbf{L}_j\star \overline{\mathcal{S}}=\overline{\mathcal{S}}\bigcup \left[
\bigcup_{j=2}^{2^{2k}}\mathbf{L}_j\star \overline{\mathcal{S}}\right].
\end{equation}

Based on our definition of the effective centralizer, it is easy to see that the group $\mathcal{\overline{G}}_N$ also has a coset structure \cite{reviewPat}. Knowing that $|\mathcal{\overline{G}}_N|=2^{2N}$ and that $|\overline{\mathcal{Z}}(\overline{\mathcal{S}})/\overline{\mathcal{S}}|=2^{N+k}$, then we can partition the effective Pauli group into cosets of the effective centralizer as 

\begin{align}  \label{pauli-centra}
\begin{split}
\overline{\mathcal{G}}_N &=\bigcup_{i=1}^{2^{N-k}}\mathbf{T}_i\star \overline{\mathcal{Z}}(\overline{\mathcal{S}})\\&=\overline{\mathcal{Z}}(\overline{\mathcal{S}})\bigcup \left[\bigcup_{i=2}^{2^{N-k}}
\mathbf{T}_i\star \overline{\mathcal{Z}}(\overline{\mathcal{S}})\right].
\end{split}
\end{align}

Equation \eqref{pauli-centra} shows how $\mathcal{\overline{G}}_N$ can be partitioned into $2^{N-k}$ cosets of the effective centralizer, each of these centralizer cosets being indexed by a coset representative\footnote{These coset representatives are commonly referred to as pure error operators in the literature.} $\{\mathbf{T}_i\}_{i=1}^{2^{N-k}}$, which when multiplied in terms of the $\star$ product by all the elements of \centra yields the specific coset in question, $\mathbf{T}_i\star\mathcal{\overline{Z}(\overline{S})}$. Finally, we can combine the partitions of \eqref{centra-cosets} and \eqref{pauli-centra}, respectively, to partition the entire effective Pauli space into cosets of the stabilizer, each one indexed by a unique $\star$ product of the $\{\mathbf{L}_j\}_{j=2}^{2^{2k}}$ and $\{\mathbf{T}_i\}_{i=1}^{2^{N-k}}$ representatives. This can be written as

\begin{align}  \label{pauli-centra_2}
\begin{split}
\overline{\mathcal{G}}_N &=\bigcup_{i=1}^{2^{N-k}}\mathbf{T}_i\star \overline{\mathcal{Z}}(\overline{\mathcal{S}})=\bigcup_{i=1}^{2^{N-k}}\bigcup_{j=1}^{2^{2k}}(\mathbf{T}_i\star \mathbf{L}_j)\star \mathcal{\overline{S}}.
\end{split}
\end{align}

\subsection{Quantum Syndromes}

In order to operate properly, QEC strategies require information regarding the transmitted quantum state. Although the axioms of quantum mechanics establish that direct measurement of a quantum state will result in information loss, quantum error syndrome measurements (Hadamard Test) can be used to measure quantum states indirectly in order to obtain sufficient information for quantum codes to operate \cite{NielsenChuang,josurev,classicaltoquantum}. A quantum syndrome $\mathrm{w} \in \mathbb{F}_2^{N-k}$ captures the commutation properties of any error sequence $\mathbf{E}\in\overline{\mathcal{G}}_N$ induced by the Pauli channel with regard to the stabilizer generators $\{\mathbf{S}_v\}_{v=1}^{N-k}$ of the code. This can be written as

\begin{equation}\label{eq:synd}
\mathrm{w}=\mathbf{e}\odot (\mathbf{s}_1,\ldots,\mathbf{s}_{N-k}),
\end{equation}

\noindent where $\mathbf{e} = \beta(\mathbf{E})$ and $\{\mathbf{s}_v\}_{v=1}^{N-k} = \{\beta(\mathbf{S}_v)\}_{v=1}^{N-k}$, respectively.
From \eqref{eq:synd} we can discern that the syndrome is a length $N-k$ binary vector whose components express the commutation relations with regard to the $\cdot$ product between $\mathbf{E}$ and each of the generators of the stabilizer code. In other words, the entries of $\mathrm{w}$ will be $0$ if the specific error operator and stabilizer generator commute and $1$ if they do not. Recall that the elements of $\mathcal{\overline{Z}(\overline{S})}$ commute with the stabilizer generators, which means that all the elements of the same effective centralizer coset will be associated to the same syndrome. For instance, if a given error pattern $\mathbf{E}'$ is associated to syndrome $\mathrm{w}_1$, all the effective Pauli sequences that belong to $\mathbf{E}'\star\mathcal{\overline{Z}(\overline{S})}$ will also be associated to syndrome $\mathrm{w}_1$. Thus, the information provided by the syndrome $\mathrm{w}$ can be used to locate the centralizer coset $\mathbf{T}_i\star\mathcal{\overline{Z}(\overline{S})}$ of a given error operator $\mathbf{E}$, where $i=1,\ldots,2^{N-k}$. Note, however, that we have no information regarding which stabilizer coset the error operator belongs to. The syndrome locates the centralizer coset that contains the error sequence that has taken place, but because all the elements of this coset are associated to the same syndrome, there is no way of knowing which specific stabilizer coset $(\mathbf{T}_i\star \mathbf{L}_j)\star \mathcal{\overline{S}}$ contains the error sequence, where $j=1,\ldots,2^{2k}$. As mentioned previously, knowing which particular stabilizer coset contains the error that has taken place is important, since operators that belong to different stabilizer cosets act non-trivially on the encoded quantum information.

\subsection{Estimation and end-to-end errors}

The effective Pauli operators induced by quantum channels and estimated by decoders can be classified into three different groups (summarized in Table \ref{errors}). Consider a channel error $\mathbf{E} \in \mathbf{T}_i\star\mathbf{L}_j\star\mathcal{\overline{S}}$ and an estimated error sequence $\mathbf{\hat{E}} \in \mathbf{\hat{T}}_i\star \mathbf{\hat{L}}_j \star \mathcal{\overline{S}}$, then the following situations can be encountered:

\begin{table*}[t!]

    \centering
    
    \caption{Characteristics of the different types of end-to-end errors that can arise when using stabilizer codes. The hat notation is used to represent estimations made by the decoder, i.e., $\hat{\mathrm{w}}$, $\hat{\mathbf{E}}$, $\mathbf{\hat{T}}_i$, and $\mathbf{\hat{L}}_j$ represent the estimated syndrome, estimated error sequence, centralizer coset representative of the estimated error sequence and stabilizer coset representative of the estimated error sequence, respectively. }
    \vspace{0.15mm}\begin{tabular}{cccc}
    \toprule 
    Type of &Defining  &Outcome &Outcome\\ error &Characteristics &(Phys. Error Rate) &(Logical Error Rate)\\
    \midrule
    End-to-end error &$\hat{\mathrm{w}} \neq \mathrm{w}$\\
     with &$\mathbf{\hat{T}}_i\neq \mathbf{T}_i$ &Failure &Failure \\
     different syndrome &$\hat{\mathbf{E}}\neq \mathbf{E}$ \\
     \midrule
    End-to-end &$\hat{\mathrm{w}} = \mathrm{w}$\\
    identical &$\mathbf{\hat{T}}_i = \mathbf{T}_i$ and $\mathbf{\hat{L}}_j\neq \mathbf{L}_j$  &Failure &Failure\\
     syndrome error &$\hat{\mathbf{E}}\neq \mathbf{E}$ \\
     \midrule
    End-to-end  &$\hat{\mathrm{w}} = \mathrm{w}$\\
    degenerate &$\mathbf{\hat{T}}_i = \mathbf{T}_i$ and $\mathbf{\hat{L}}_j = \mathbf{L}_j$  &Failure &Success\\
    error  &$\hat{\mathbf{E}}\neq \mathbf{E}$ \\
    \bottomrule
    \label{errors}
    \end{tabular}
    \end{table*}

\begin{itemize}

 \item \textbf{End-to-end errors with different syndromes:}  These events occur when the error sequence estimated by the decoder and the real error sequence belong to different centralizer cosets, i.e., $\mathbf{T}_i\neq\mathbf{\hat{T}}_i$. Equipped with an ideal decoder, such scenarios would not exist, as the syndrome of the estimated error pattern $\hat{\mathbf{E}}$ should always match the measured syndrome associated to the channel error $\mathbf{E}$. However, because sparse quantum codes are generally decoded based on the sub-optimal classical SPA algorithm\footnote{The SPA algorithm solves the SWML (Symbol-Wise Maximum Likelihood) problem and not the actual ML (Maximum Likelihood) decoding problem. While the solutions to the SWML and ML problems usually coincide, this is not always the case.}, these errors can take place with varying probability \cite{degen3, reviewPat, mod-BP, Pat-thesis}. 
 
\item \textbf{End-to-end identical syndrome errors:} These events take place when the estimated error sequence and the channel error both belong to the same centralizer coset, $\mathbf{T}_i = \mathbf{\hat{T}}_i$, but each of them belongs to a different stabilizer coset, i.e., $\mathbf{L}_j \neq \mathbf{\hat{L}}_j$ (the channel logical operator and estimated logical operator do not match). Thus, although $\hat{\mathbf{E}}$ and $\mathbf{E}$ exhibit identical commutation properties with respect to the stabilizer generators, they will each act on the transmitted codeword in a distinct non-trivial manner and the decoder will fail. It should be noted that in the literature this type of end-to-end error is generally referred to as a logical error.

\item \textbf{End-to-end degenerate errors:} These events take place when the estimated error pattern and the channel error both belong to the same stabilizer coset, $\mathbf{T}_i = \mathbf{\hat{T}}_i$ and $\mathbf{L}_j = \mathbf{\hat{L}}_j$, but they do not match, i.e., $\hat{\mathbf{E}} \neq \mathbf{E}$. Since the estimated error belongs to the same stabilizer coset as the channel error, it will act identically on the quantum codeword, which means that it will not actually result in a decoding failure and should not be considered as such. 
\end{itemize}

\section{Performance assessment metrics}

The manifestation of the degeneracy phenomenon in the realm of quantum error correction makes it impossible to accurately predict the performance of QEC codes with methods that disregard its presence. Thus, in the paradigm of sparse quantum codes, which are famous for their degenerate nature, appropriate performance assessment is of paramount importance. This can be achieved using the logical error rate, which accurately predicts the performance of quantum codes and distinguishes between the different types of end-to-end errors (see table \ref{errors}). 

Prior to discussing how to compute the logical error rate, it should be mentioned that the concept of \textit{undetected} errors is not exclusive to the quantum paradigm. In fact, even though degeneracy does not exist in the classical coding framework, \textit{undetected} or \textit{logical} errors in classical LDPC codes and classical turbo codes have previously been studied. This idea was introduced by the early work of MacKay et al. \cite{Mac}, where a classical undetected error is defined as a decoding estimate that is not equal to the original error sequence and that is produced when the decoder exits before the maximum number of decoding iterations (it produces a valid syndrome). In their analysis of classical LDPC codes, MacKay et al. showed that all of the decoding mistakes they encountered were detected errors (classical undetected errors were only observed in turbo codes). This was also shown in \cite{refined} for a slightly different decoding algorithm. Similar outcomes were observed in the quantum paradigm for the failed recoveries of the modified decoding strategies of \cite{degen3, efb}. These techniques serve to improve standard SPA decoding of quantum codes by post-processing the initial error estimates and producing new estimates of the channel error. If these new estimates do not revert the channel error, then they are referred to as failed recoveries or failed error corrections. In \cite{degen3}, all of the failed error corrections were shown to be end-to-end errors with different syndromes, whereas in \cite{efb} a small percentage of failed estimates were shown to be end-to-end identical syndrome errors and end-to-end degenerate errors\footnote{The authors of this work do not distinguish between end-to-end identical syndrome errors and end-to-end degenerate errors.}.

Aside from these failed recovery analyses, the literature is limited when it comes to assessing the percentage of decoding failures that is caused by each type of end-to-end error. It is reasonable to believe that QLDPC codes, given their large number of degenerate operators \cite{degen3, degen4, Hard, reviewPat, QEClidar, trapping-sets}, will experience a large percentage of end-to-end degenerate errors. We confirm this intuition in the final section of our work, where we show how up to $30\%$ of the end-to-end errors that take place when using the QLDPC codes of \cite{jgf1, jgf2, patrick1} are degenerate. 

\subsection{Discriminating between different types of end-to-end errors}

It is clear from the defining characteristics of each specific type of end-to-end error (see table \ref{errors}), that \textit{end-to-end errors with different syndromes} are the easiest type of error to identify. In fact, doing so is trivial, as all that is required is a comparison of the syndrome estimate, $\mathrm{\hat{w}} \in  \mathbb{F}_2^{N-k}$, and the measured syndrome, $\mathrm{w} \in \mathbb{F}_2^{N-k}$, where $k$ and $N$ represent the number of logical qubits and physical qubits (blocklength) of the quantum code, respectively. Similarly, knowing that either an \textit{end-to-end identical syndrome error} or an \textit{end-to-end degenerate error} has occurred is simple. This can be done by comparing the estimate of the error sequence $\mathbf{\hat{E}}$ to the channel error $\mathbf{E}$ whenever $\mathrm{\hat{w}} = \mathrm{w}$, i.e., if $\mathbf{\hat{E}} \neq \mathbf{E}$ either an end-to-end degenerate error or an identical syndrome error has occurred, and if $\mathbf{\hat{E}} = \mathbf{E}$, no error has taken place. The issue arises when trying to distinguish between these two families of end-to-end errors. Notice that the comparison $\mathbf{\hat{E}} \neq \mathbf{E}$ does not reveal if the error estimate belongs to the same stabilizer coset as the channel error. Hence, we have no way of discriminating between end-to-end degenerate errors and end-to-end identical syndrome errors.

Conceptually, the simplest and most straightforward strategy that comes to mind to resolve this problem is to compute the stabilizer of the code in question,  compute the $\star$ product of the stabilizer with the channel error $\mathbf{E}$ to extract the specific stabilizer coset of the channel error, and then check if $\mathbf{\hat{E}}$ belongs to this coset. As mentioned in \cite{reviewPat}, this works because the coset representative choice for the coset $(\mathbf{T}_i\star\mathbf{L}_j)\star\mathcal{\overline{S}}$ is irrelevant (any operator belonging to the coset serves as a valid representative), where $\{\mathbf{T}_i, \mathbf{L}_j\} \in \mathcal{\overline{G}}_N$, and $\mathcal{\overline{S}} \subset \mathcal{\overline{G}}_N$, represent the effective centralizer coset representative, the stabilizer coset representative, and the stabilizer itself. Thus, computing $\mathbf{E} \star \mathcal{\overline{S}}$ will yield the stabilizer coset of the channel error, i.e.,  $\mathbf{E} \star \mathcal{\overline{S}} = (\mathbf{T}_i\star\mathbf{L}_j)\star\mathcal{\overline{S}}$. Therefore, whenever $\mathrm{\hat{w}} = \mathrm{w}$ and $\mathbf{\hat{E}} \neq \mathbf{E}$, we will know that an end-to-end degenerate error has occurred if the estimated error sequence $\mathbf{\hat{E}}$ is in the coset $\mathbf{E} \star \mathcal{\overline{S}}$. If this does not occur, then an end-to-end identical syndrome error will have taken place. 

Unfortunately, since extracting the stabilizer of a quantum code becomes increasingly complex as its blocklength increases, this strategy will only be applicable to short quantum codes. The number of elements in the stabilizer of a quantum code with blocklength $N$ and rate $R_Q$ is given by $2^{N-k}=2^{N(1-R_Q)}$ \cite{reviewPat}, which grows exponentially with $N$ and can rapidly become intractable on a classical machine as this parameter increases. In light of this, it is apparent that more practical methods to differentiate between end-to-end degenerate errors and end-to-end identical syndrome errors are necessary. Both the strategy we propose herein and that employed in \cite{degen3, refined, osd1, osd2, exploiting}, which are explained in the next two sections, resolve this issue.

\subsection{An algebraic perspective on end-to-end degenerate errors}

The problem of differentiating between end-to-end identical syndrome errors and end-to-end degenerate errors can also be formulated as a set of linear equations. The Parity Check Matrix (PCM) of a stabilizer code that encodes $k$ logical qubits into $N$ physical qubits (a rate $R_Q = \frac{k}{N}$ code with blocklength $N$) can be written as
\begin{equation}\label{eq:pcm}
\mathbf{H}_{\mathcal{\overline{S}}} = \begin{pmatrix}
\mathbf{h}_1 \\
\mathbf{h}_2 \\
\vdots \\
\mathbf{h}_{N-k}
\end{pmatrix},
\end{equation}
where $\mathbf{h}_v = \mathbf{s}_v$ denotes the symplectic representation of the generators $\{\mathbf{S}_v\}_{v=1}^{N-k} \in \mathcal{\overline{G}}_N$ that define the stabilizer group $\mathcal{\overline{S}}$. Each of the elements of $\mathcal{\overline{S}}$ is a linear combination of the $N-k$ generators, hence, if $\mathbf{S}$ is an element of the stabilizer and $\mathbf{s}$ is the symplectic representation of this stabilizer element, then
\begin{equation}\label{eq:stabsarelinearcombs}
\mathbf{s} = \left(\sum_{i=1}^{N-k} \mathrm{a}_i \mathbf{h}_i\right)\mod{2},
\end{equation}
where $\mathbf{a} = (\mathrm{a}_1,\ldots,\mathrm{a}_{N-k})$ is a unique binary vector.

Whenever a channel error $\mathbf{E}$ takes place, the decoder will compute an estimate of this error and produce an estimate of the syndrome associated to it. As discussed previously, this syndrome only determines which specific effective centralizer coset the channel error belongs to. In other words, the syndrome provides the effective centralizer coset representative $\mathbf{T}_i$ (known as the pure error component in the literature) of the channel error. Assuming that an end-to-end error with different syndrome does not occur, the estimated syndrome will match the measured syndrome, hence the centralizer coset representative of the estimated error sequence and the centralizer coset representative of the channel error will also be the same, i.e., $\mathbf{T}_i = \mathbf{\hat{T}}_i$. Thus, if we compute the $\star$ operation of the channel error $\mathbf{E}$ and the estimated error $\mathbf{\hat{E}}$, which can also be understood as the mod2 sum of their symplectic representations over $\mathbb{F}_2^{2N}$: $\beta(\mathbf{E})\oplus \beta(\mathbf{\hat{E}}) = \mathbf{e}\oplus\mathbf{\hat{e}}$, where $\beta$ denotes the symplectic map, the sequence $\mathbf{E}$ will be shifted to the effective centralizer $\mathcal{\overline{Z}}(\mathcal{\overline{S}})$. Note that we can also write this using the symplectic map as $\mathbf{E}\star\mathbf{\hat{E}} \in \mathcal{\overline{Z}}(\mathcal{\overline{S}}) \rightarrow \beta(\mathbf{E})\oplus \beta(\mathbf{\hat{E}}) \in \beta(\mathcal{\overline{Z}}(\mathcal{\overline{S}})) = \mathbf{e}\oplus\mathbf{\hat{e}}\in\Gamma_R$, where $\Gamma_R \subset \mathbb{F}_2^{2N}$ denotes the equivalent group of the effective centralizer over $\mathbb{F}_2^{2N}$. Based on this, the issue of determining whether an end-to-end error is degenerate can be understood as finding out if $\mathbf{E}\star\mathbf{\hat{E}}$ belongs to the stabilizer. This can be formulated based on the symplectic map into the following question:
\begin{equation}\label{eq:question}
\exists \ \mathbf{a} : \mathbf{e}\oplus\mathbf{\hat{e}} = \left(\sum_{i=1}^{N-k}\mathrm{a}_i \mathbf{h}_i\right)\mod{2} \ \text{ ?}
\end{equation}

Essentially, if a set of coefficients $\mathbf{a}$ exists such that the above equation holds, i.e., if $\mathbf{e}\oplus\mathbf{\hat{e}} \in \mathbb{F}_2^{2N}$ is a linear combination of the symplectic representation of the stabilizer generators, then $\mathbf{E}\star\mathbf{\hat{E}}$ belongs to the stabilizer and an end-to-end degenerate error will have occurred. If such a set of coefficients does not exist, then and end-to-end identical syndrome error has taken place.

The expression shown in \eqref{eq:question} defines a linear system of equations over the binary field. An answer to this question can be found by writing the augmented matrix $[\mathbf{H}_{\mathcal{\overline{S}}}^\top | (\mathbf{e}\oplus\mathbf{\hat{e}})^\top]$ in its row-echelon form. This means that, based on this procedure, it is possible to determine the type of end-to-end errors that occur and subsequently compute the logical error rate. This is done in \cite{mod-BP}. However, although the procedure is conceptually simple, rewriting the augmented matrix in such a manner becomes increasingly computationally complex as matrices grow in size. Unfortunately, for sparse quantum codes to be good, the blocklength must be large, which implies that the PCMs of these codes will also be large\footnote{The PCMs of QEC codes are of size $N-k\times 2N$. A common size in the literature of QLDPC codes is $10000$ qubits, thus the PCM associated to such a code would be of size $(10000-k)\times 20000$.}. Furthermore, the row-echelon form of the matrix $[\mathbf{H}_{\mathcal{\overline{S}}}^\top | (\mathbf{e}\oplus\mathbf{\hat{e}})^\top]$ must be computed during every simulation iteration (whenever the estimated syndrome and the measured syndrome match) to determine what type of end-to-end error has taken place, which may significantly increase simulation time. For these reasons, calculating the logical error rate based on this procedure can become a cumbersome and lengthy endeavour. Therefore, the task at hand is to find a more practical and less computationally demanding way to determine if the congruence equation system given in \eqref{eq:question} has a solution, as this suffices to determine if the end-to-end error is degenerate (we do not actually need to solve the system itself).

\subsubsection{Classical coding-inspired strategy}

It is possible to find an answer to \eqref{eq:question} by casting the problem in the framework of classical linear block codes. In classical coding theory, the encoding matrix or generator matrix $\mathbf{G}_c$ of a binary linear block code and its corresponding parity check matrix $\mathbf{H}_c$ fulfil $(\mathbf{G}_c\mathbf{H}_c^\top)\mod2 = (\mathbf{H}_c\mathbf{G}_c^\top)\mod2 = 0$. This means that the parity check matrix defines a basis for the nullspace of the generator matrix and viceversa. In the classical scenario, having a basis for the nullspace of a code enables us to determine whether the decoding outcome $\mathbf{x}$ belongs to the code by simply computing its product with the parity check matrix of the code, i.e., if $(\mathbf{H}_c\mathbf{x}^\top)\mod 2 = 0$ then $\mathbf{x}$ is a codeword. Essentially, whenever $(\mathbf{H}_c\mathbf{x}^\top)\mod 2 = 0$, the decoding outcome is a linear combination of the rows of the generator matrix $\mathbf{G}_c$ and it belongs to the code, and whenever $(\mathbf{H}_c\mathbf{x}^\top)\mod 2 \neq 0$, $\mathbf{x}$ does not belong to the code. 

Notice that, based on this formulation, the quandary posed in \eqref{eq:question} is reminiscent of the classical decoding scenario. The main difference is that instead of determining if the decoding outcome belongs to the code, we must discover if the sum of the symplectic representations of the channel error and the estimated error belong to the stabilizer. This parallelism between the classical and quantum problems allows us to apply the classical resolution strategy to the quantum paradigm with only a slight caveat: answering \eqref{eq:question} requires an inverse approach to the classical method. Since the generators of the stabilizer code are given by the rows of the parity check matrix $\mathbf{H}_{\mathcal{\overline{S}}}$, the corresponding kernel generator matrix\footnote{We refer to the matrix $\mathbf{G}_{\mathcal{\overline{S}}}$ as the kernel generator matrix to avoid the term stabilizer generator matrix, as this latter term implies that the the matrix can be used for encoding purposes (which may not be true in the present case).} $\mathbf{G}_{\mathcal{\overline{S}}}$ (instead of the parity check matrix like in the classical paradigm) must be used to discover if $\mathbf{e}\oplus\mathbf{\hat{e}}$ can be written as a linear combination of the stabilizer generators. The matrix $\mathbf{G}_{\mathcal{\overline{S}}}$ defines a basis for the nullspace of the stabilizer code, hence it will suffice to compute $[\mathbf{G}_{\mathcal{\overline{S}}}(\mathbf{e}\oplus\mathbf{\hat{e}})^\top]\mod 2$ to find the answer to \eqref{eq:question}. If $[\mathbf{G}_{\mathcal{\overline{S}}}(\mathbf{e}\oplus\mathbf{\hat{e}})^\top]\mod 2 = 0$, $\mathbf{E}\star\mathbf{\hat{E}} \in \mathcal{\overline{S}}$ which means that an end-to-end degenerate error has occurred, and if $[\mathbf{G}_{\mathcal{\overline{S}}}(\mathbf{e}\oplus\mathbf{\hat{e}})^\top]\mod2 \neq 0$, $\mathbf{E}\star\mathbf{\hat{E}} \not\in\mathcal{\overline{S}}$, and an end-to-end identical syndrome error will have taken place.

This strategy provides us with a simple and computationally efficient method to determine the type of end-to-end error that has taken place. The only requirement is obtaining the matrix $\mathbf{G}_{\mathcal{\overline{S}}}$, which can be computed once (by finding a basis for the nullspace of its parity check matrix $\mathbf{H}_{\mathcal{\overline{S}}}$) and can then be stored offline for any stabilizer code. In this manner, we have designed a simple method to solve \eqref{eq:question} that does not require the computation of the stabilizer and so avoids the complexity issues that this entails. 

\subsection{ Detecting end-to-end degenerate errors using encoded Pauli operators}

 There is another manner of distinguishing between end-to-end identical syndrome errors and end-to-end degenerate errors. It involves obtaining the encoded Pauli operators\footnote{In the literature, these operators are referred to as logical operators. However, we use the term encoded Pauli operators to distinguish them from the stabilizer coset representatives $\{\mathbf{L}_j\}_{j=1}^{2^{2k}}$ which we originally defined as logical operators.} of a code following a method derived by Gottesman in his seminal work \cite{QSC}, and then using these operators to determine whether the error estimate produced by the decoder is in the stabilizer coset of the channel error (end-to-end degenerate errors occur when this happens and end-to-end identical syndrome errors occur when it does not). This strategy was first applied to QLDPC codes in \cite{degen3}, and has since been used in \cite{refined, osd1, osd2, exploiting}. 
 
  The encoded Pauli operators of an $N$-qubit stabilizer code are defined as those operators in $\mathcal{G}_N$ that commute with the elements of the stabilizer group and whose action on an encoded state can be understood as an $X, Y, $ or $Z$ operation on each of the encoded logical qubits. Each stabilizer code has $2k$ encoded Pauli operators\footnote{Given the difference in the number of stabilizer coset representatives (logical operators) $\{\mathbf{L}_j\}_{j=1}^{2^{2k}}$ and encoded Pauli operators, it is useful to employ different terminology for these concepts (even though encoded Pauli operators are related to logical operators and vice-versa).} which are generally represented using the notation $\overline{Z}_q$ and $\overline{X}_l$, where $\{q,l\}\in\{1,\ldots,k\}$. In this manner, $\overline{Z}_q$ represents an operator in $\mathcal{G}_N$ whose action is analogous to performing a $Z$ operation (phase flip) on the $i$-th logical qubit. Thus $\overline{Z}_q \in \mathcal{G}_N$ maps to $Z_q \in \mathcal{G}_k$, where $Z_q$ denotes the action of a $Z$ operator on the $i$-th qubit and the action of $I$ operators on the remaining qubits (identity operators are omitted). Recall that, since the global phase can be ignored, it will be equivalent for the purpose of error correction to consider the equivalent encoded Pauli operators over the effective $N$-fold Pauli group, $\overline{Z}_q \equiv \overline{\mathbf{Z}}_q$ and $\overline{X}_l \equiv \overline{\mathbf{X}}_l$, where $\overline{\mathbf{Z}}_q, \overline{\mathbf{X}}_l \in \mathcal{\overline{G}}_N$ (we use capital boldface to preserve the notation of \cite{reviewPat}.)  
  
  Based on this definition of the encoded Pauli operators, we know that an encoded Pauli operator $\overline{\mathbf{Z}}_q$ commutes with all the elements of $\mathcal{\overline{S}}$ as well as with all other encoded Pauli operators except for the operator $\overline{\mathbf{X}}_l$ when $q=l$. This means that, if the encoded Pauli operators of a stabilizer code are known, we can determine if an operator $\mathbf{A} \in \mathcal{\overline{Z}(\overline{S})} \subset \mathcal{\overline{G}}_N$ belongs to $\mathcal{\overline{S}}$ by checking the commutation relations of $\mathbf{A}$ with the encoded Pauli operators. If $\mathbf{A}$ commutes with all the encoded Pauli operators it is within the stabilizer and if it does not (it anti commutes with one encoded Pauli operator) it is not in the stabilizer. Against this backdrop, it is easy to see how this strategy can be applied to solve the issue of discriminating between end-to-end identical syndrome errors and end-to-end degenerate errors. After successful decoding (the decoder produces a matching estimate of the syndrome), we know that $\mathbf{E}\star\mathbf{\hat{E}} \in\mathcal{\overline{Z}(\overline{S})}$. Now, we determine if  $\mathbf{E}\star\mathbf{\hat{E}}$ is in  $\mathcal{\overline{S}}$ by checking its commutation status with $\{\overline{\mathbf{Z}}_q\}_{q=1}^k$ and $\{\overline{\mathbf{X}}_l\}_{l=1}^k$. If $\mathbf{E}\star\mathbf{\hat{E}}$ commutes with all the encoded Pauli operators, $\mathbf{E}\star\mathbf{\hat{E}} \in \mathcal{\overline{S}}$ and an end-to-end degenerate error has occurred. If not, an end-to-end identical syndrome error has occurred. It is based on these comparisons that the logical error rate was successfully computed in \cite{osd1, osd2, refined, exploiting}.
  
  Naturally, to be able to apply the method one must first have knowledge of the encoded Pauli operators of the code. This is similar to the classical coding-based strategy we propose in this paper, which requires the computation of the kernel generator matrix $\mathbf{G}_{\mathcal{\overline{S}}}$. The encoded Pauli operators of a stabilizer code can be found based on the concept of the \textit{standard form} (see Chapter 4 of \cite{QSC}). In this work, Gottesman showed how, by applying row operations (i.e. Gaussian elimination) together with the necessary qubit permutations (i.e. column permutations) on the parity check matrix of a stabilizer code, one can obtain a special row reduced echelon form of the parity check matrix: the standard form. Once the standard form is known, the encoded Pauli operators $\{\overline{\mathbf{Z}}_q\}_{q=1}^k$ and $\{\overline{\mathbf{X}}_l\}_{l=1}^k$ can be directly obtained from it using matrix algebra \cite{QSC, Benny}. 
  As with the kernel generator matrix $\mathbf{G}_{\mathcal{\overline{S}}}$, the encoded Pauli operators of a specific code need only be computed once and can then be stored offline. 
  
   \subsection{Method Comparison}
   
    For general stabilizer codes, computing the logical error rate based on the classical coding inspired strategy does not yield benefits over doing so via the encoded Pauli operator method. This has to do with the fact that both of them have the same computational complexity: performing Gaussian elimination and column permutations is also what finding the basis for the nullspace of a matrix requires. As is shown in \cite{complex}, these operations scale as $O(N^3)$.
   However, when it comes to the widely-employed family of Calderbank-Shor-Steane codes, the classical coding inspired strategy we propose herein will benefit from a substantial complexity advantage.
   
   CSS codes \cite{CSS2, CSS1} are a particular type of stabilizer code that provide an efficient design strategy to build QEC codes from existing classical codes. The quantum parity check matrix of a CSS code is written as

\begin{equation} \label{eq:CSS}
\mathbf{H}_Q = (\mathrm{H}_x|\mathrm{H}_z) =
\begin{pmatrix} \mathrm{H}_x^{'} &0 \\
    0 &\mathrm{H}_z^{'} \end{pmatrix} , 
\end{equation}

where $\mathrm{H}_x = \begin{pmatrix} \mathrm{H}_x^{'} \\ 0 \end{pmatrix}$ and $\mathrm{H}_z = \begin{pmatrix} 0 \\ \mathrm{H}_z^{'} \end{pmatrix}$.

In this construction, $\mathrm{H}_x'$ and $\mathrm{H}_z'$ are the parity check matrices of two binary classical LDPC codes $C_1$ and $C_2$, respectively, where each matrix is used to correct either bit-flips or phase-flips. As discussed previously, any classical binary LDPC code or any classical binary linear block code for that matter, will satisfy $(\mathbf{G}_c\mathbf{H}_c^\top)\text{mod}2 = \mathbf{0}$, where $\mathbf{G}_c$ and $\mathbf{H}_c$ represent the generator and parity check matrices of a binary classical LDPC code, respectively. Because the generator matrix is necessary to encode a classical LDPC code, the issue of deriving the matrix $\mathbf{G}_c$ from $\mathbf{H}_c$ has been widely researched in the realm of classical error correction \cite{complex, G1,G2,G3,G4}. In general, finding the matrix $\mathbf{G}_c$ from $\mathbf{H}_c$, i.e, bringing the PCM into the desired form, requires $O(N^3)$ operations of
processing \cite{complex}. However, through the clever application of optimization algorithms, the complexity of this task has been reduced \cite{complex, G1, G2} and classical LDPC codes with a linear encoding complexity\footnote{Richardson et al.\cite{complex} showed that the encoding complexity is upper-bounded by $O(N) + O(g^2)$, where $g$ is the gap to measure the ``distance'' between a given parity-check matrix and a lower triangular matrix. This means that in the extreme case where $g=0$, such as for Irregular Repeat–Accumulate (IRA) codes \cite{IRA}, $\mathbf{G}_c$ can be computed with complexity $O(N)$.} have been found. Note that based on the structure of the QPCM of CSS codes \eqref{eq:CSS} we can derive the kernel generator matrix of a CSS code as


\begin{equation} \label{eq:CSS2}
\mathbf{G}_\mathcal{\overline{S}} = (\mathrm{G}_x|\mathrm{G}_z) =
\begin{pmatrix} \mathrm{G}_x^{'} &0 \\
    0 &\mathrm{G}_z^{'} \end{pmatrix} , 
\end{equation}

where $\mathrm{G}_x'$ and $\mathrm{G}_z'$ are the generator matrices of the binary classical LDPC codes $C_1$ and $C_2$, respectively. We know that the matrix given in \eqref{eq:CSS2} will actually serve the purpose of a kernel generator matrix because its rows are all linearly independent. The difficulty in computing \eqref{eq:CSS2} lies in deriving the generator matrices $\mathrm{G}_x'$ and $\mathrm{G}_z'$ from the PCMs $\mathrm{H}_x'$ and $\mathrm{H}_z'$. However, when constructing CSS QLDPC codes from classical LDPC codes, it is a prerequisite that the classical code be known. Thus, it is reasonable to assume that the generator matrices of the classical codes that make up a CSS QLDPC code are available\footnote{If the LDPC codes have previously been used in the classical paradigm then these generator matrices are known (they must have been employed in the classical encoding procedure).}. Against this backdrop, because this does not require finding the standard form of the QPCM and deriving its encoded Pauli operators, the method we propose herein will have a complexity advantage over the encoded Pauli operator method when computing the logical error rate of sparse CSS quantum codes. More specifically, when using classical LDPC codes to build QLDPC CSS codes, because the generator matrices associated to the corresponding parity check matrices of these codes are known, the complexity of our method is negligible, as no calculations will be necessary to obtain the kernel generator matrix (it can be derived by simply introducing the appropriate generator matrices in (14), i.e, the complexity is $O(1)$). In the case that the generator matrix of a particular classical LDPC code is not known, it can be derived as shown in \cite{complex} with complexity $O(N) + O(g^2)$, where $g$ is the gap to measure the ``distance'' between the parity-check matrix and a lower triangular matrix. This means that the computational complexity of calculating the kernel generator matrix of a CSS code will be at most $O(N) + O(g^2)$, which implies that, in the context of CSS codes, our strategy is less computationally complex than the one based on using encoded Pauli operators, which has complexity $O(N^3)$.

\section{Simulation Results} \label{sec:firstresults}

\begin{figure}[!htp]
\begin{minipage}[b]{\columnwidth}	
\centering
  \includegraphics[width=\linewidth, height=2.5in]{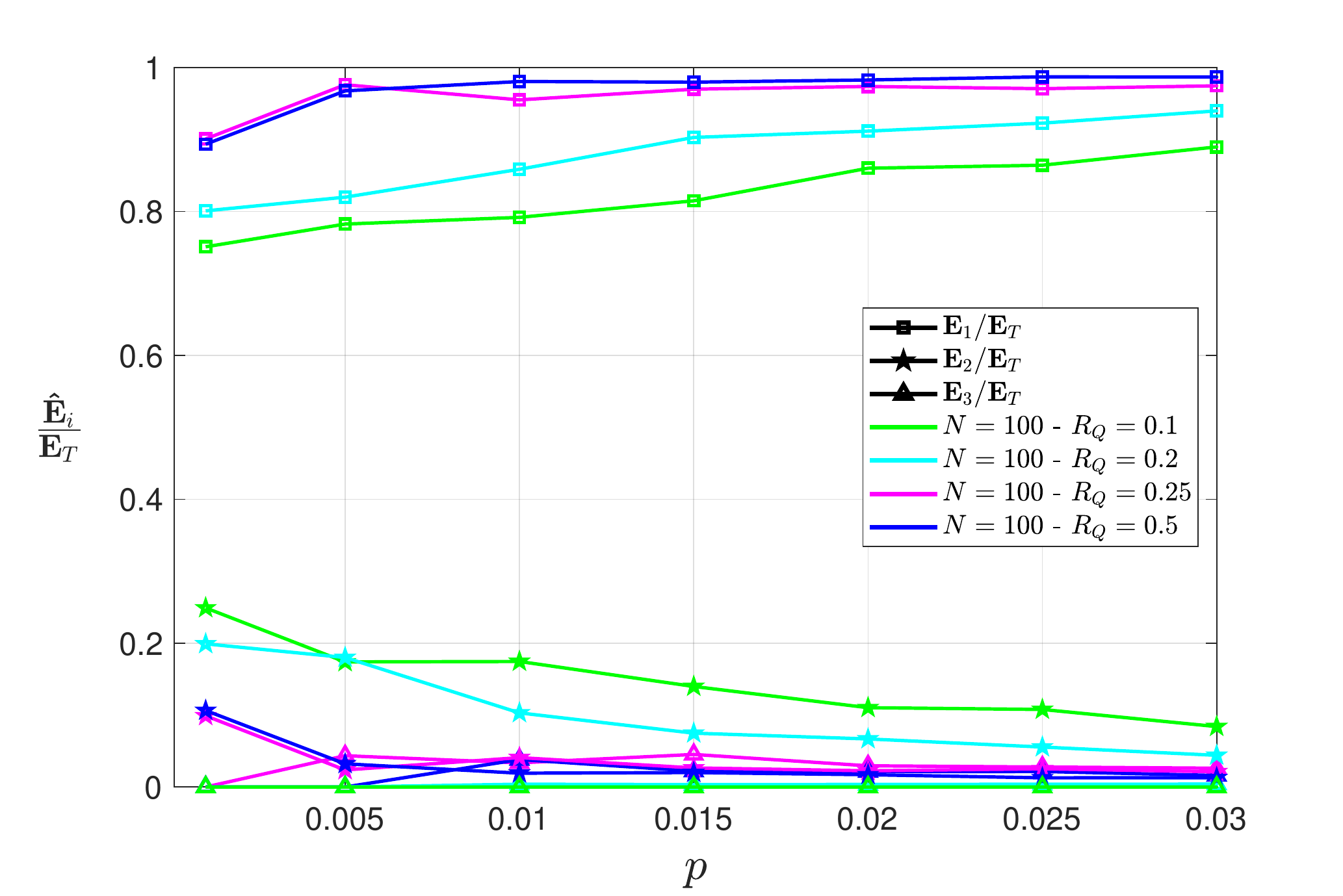}
  {\\(a)}
    \label{fig:sub11}
\end{minipage}%
\hspace{1cm}
\begin{minipage}[b]{\columnwidth}
\centering
    \includegraphics[width=\linewidth, height=2.5in]{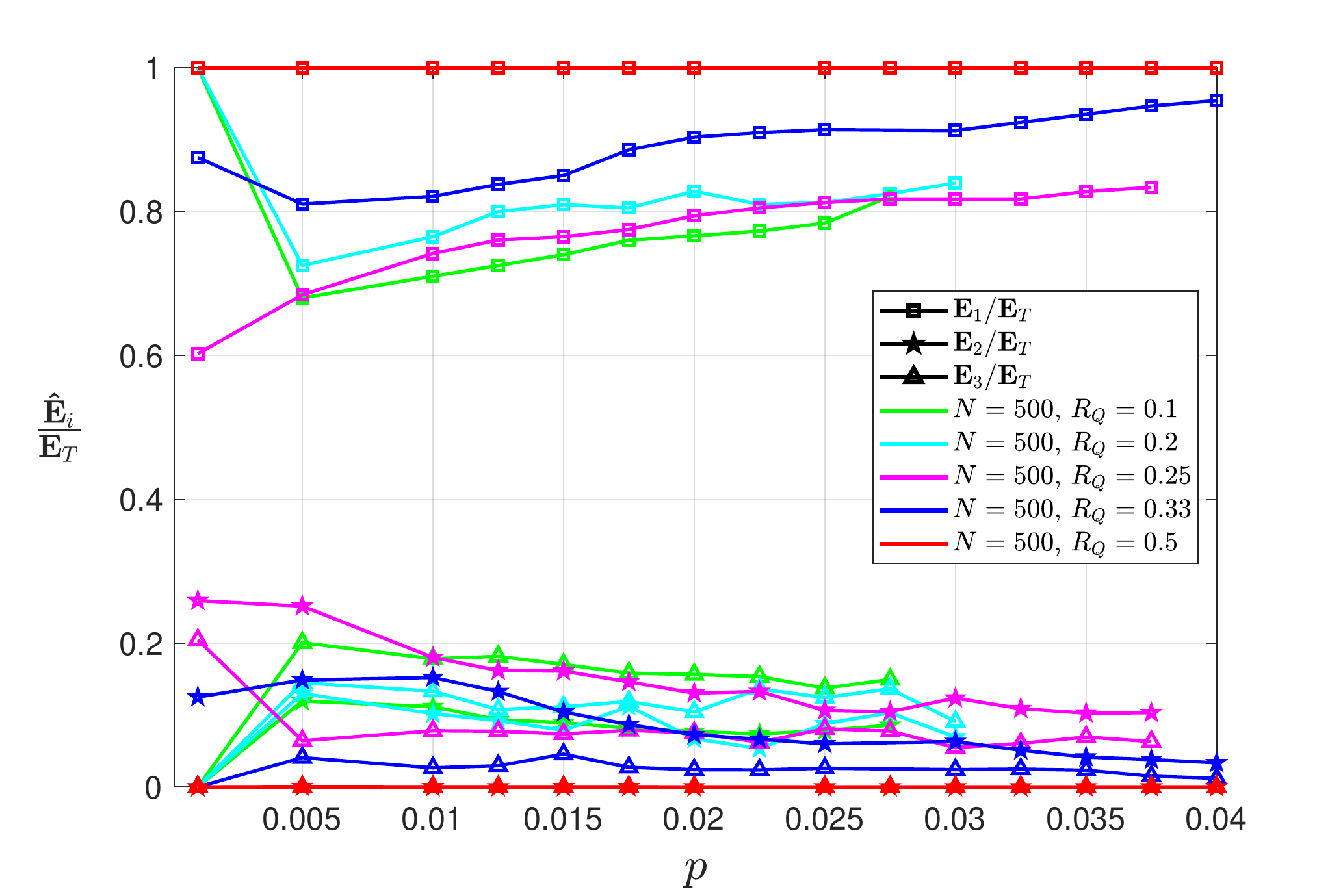}
  {\\(b)}
     \label{fig:sub12}
\end{minipage}
\begin{minipage}[b]{\columnwidth}
\centering
  \includegraphics[width=\linewidth,height=2.5in]{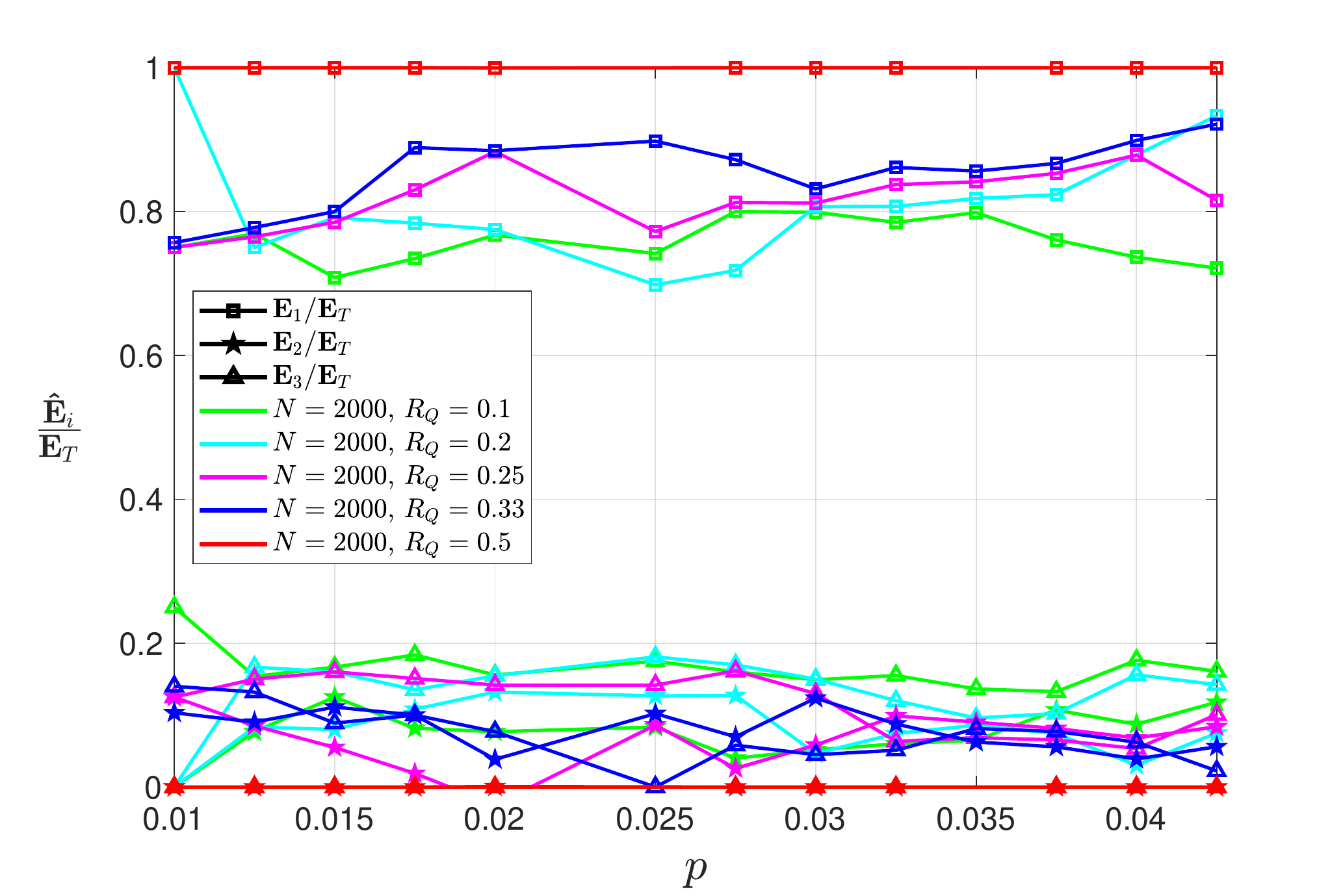}
  {\\(c)}
  \caption{Ratios of end-to-end different syndrome, identical syndrome and degenerate errors for codes of various rates and blocklengths: (a) $N=100$ (b) $N=500$ (c) $N=2000$.}
\label{fig:results-codes}
  
\end{minipage}
 \label{fig:degen-meth}
\end{figure}

We close this section by using our method to show how end-to-end errors with identical syndromes and end-to-end degenerate errors can account for a significant percentage of the end-to-end errors that occur when using QLDPC codes. For this purpose, we simulate the CSS QLDGM codes of \cite{jgf1, jgf2, patrick1} with different rates and blocklengths over the depolarizing channel. The characteristics of these codes are detailed in table \ref{tab:codes}. The underlying classical LDGM matrices and the structure of the $M$ matrix, critical to the performance of the QLDGM codes, have been chosen according to the optimization guidelines detailed in \cite{jgf1, jgf2, patrick1, patrick2, patrick3}. The results of these simulations are shown in Figure \ref{fig:results-codes}, where each subfigure groups the results by blocklength, i.e, each of the subfigures portrays the results for all the codes with the same value of $N$. The graphs plot the ratio of a specific type of end-to-end error against the depolarizing probability. The aforementioned ratio is computed as $\frac{\mathrm{E}_i}{\mathrm{E}_T}$, where $\mathrm{E}_i$ denotes the total number of end-to-end errors of a specific type ($i = 1,2,3$.): 
\begin{itemize}
    \item $\mathrm{E}_1 \rightarrow$ end-to-end errors with different syndromes,
    \item $\mathrm{E}_2 \rightarrow$ end-to-end errors with identical syndromes,
    \item $\mathrm{E}_3 \rightarrow$ end-to-end degenerate errors,
    
\end{itemize} and $\mathrm{E}_T$ represents the total number of end-to-end errors. To ensure that the simulation results are precise, the ratios have been computed after a total of $1000$ decoding mistakes have been made (following the Monte Carlo simulation rule of thumb provided in \cite{MonteCarlo}), i.e., $\mathrm{E}_T = 1000$.

\begin{table}[h!]    

    \caption{\normalsize{Parameter values and configurations of simulated codes. }}
    \centering
    \begin{tabular}{ccccc}
    \toprule 
    $N$ &$R_Q$ &Classical LDGM &$[m, p, x, y]$\\
    \midrule
    $100$ &$0.1$ &P($3,3$) &[$45, 24, 6, 3$] \\
    $100$ &$0.2$ &P($3,3$) &[$40, 18, 6, 3$] \\
    $100$ &$0.25$ &P($3,3$) &[$38, 15, 6, 3$] \\
    $100$ &$0.5$ &P($3,3$) &[$25, 6, 7.57, 3$] \\\midrule
    $500$ &$0.1$ &P($5,5$) &[$225, 170, 11, 3$] \\
    $500$ &$0.2$ &P($5,5$) &[$200, 144, 11, 3$] \\
    $500$ &$0.25$ &P($5,5$) &[$188, 130, 11, 3$] \\
    $500$ &$0.33$ &P($5,5$) &[$163, 102, 11, 3$] \\
    $500$ &$0.5$ &P($5,5$) &[$125, 60, 11, 3$] \\\midrule
    $2000$ &$0.1$ &P($9,9$) &[$900, 691, 11, 3$] \\
    $2000$ &$0.2$ &P($9,9$) &[$800, 581, 11, 3$] \\
    $2000$ &$0.25$ &P($9,9$) &[$750, 526, 11, 3$] \\
    $2000$ &$0.33$ &P($9,9$) &[$670, 438, 11, 3$] \\
    $2000$ &$0.5$ &P($9,9$) &[$500, 251, 11, 3$] \\
    \bottomrule
    \end{tabular}
    \label{tab:codes}
    \end{table}

The outcomes portrayed in Figure \ref{fig:results-codes} confirm our initial intuition that sparse quantum codes are degenerate. It is easy to see that for all of the simulated blocklengths and rates (except for $R_Q = 0.5$), the percentage of end-to-end errors that are not of the $\mathrm{E}_1$ type is not negligible, i.e., $\frac{\mathrm{E}_2}{\mathrm{E}_T} + \frac{\mathrm{E}_3}{\mathrm{E}_T} \neq 0$. Furthermore, these results speak towards the higher precision of the logical error rate compared to the physical error rate when assessing the performance of these codes. For instance, at a noise level of $p = 0.005$, $\frac{\mathrm{E}_3}{\mathrm{E}_T} = 0.198$ for the $N=500$ $R_Q = 0.1$ code. This means that $19.8\%$ of the end-to-end errors are degenerate and should not be counted as decoding failures. Thus, in this scenario, the physical error rate overestimates the number of decoding failures and does not provide an accurate representation of the performance of the code. In fact, regardless of the noise level of the channel, the rate of the code (except for $R_Q = 0.5$), and the blocklength of the code, end-to-end degenerate errors take place, and so the physical error rate will always provide an inaccurate representation of the performance of these sparse quantum codes. Therefore, as is stated in \cite{exploiting}, it is clear that performance results assessed based on the physical error rate ($\mathbf{\hat{E}} = \mathbf{E}$ as the decoding success criterion) \cite{jgf1, jgf2, qldpc15, efb, bab1, patrick1} are inaccurate (they report an upper bound).

Furthermore, the results shown in Figure \ref{fig:results-codes} also reveal how the frequency with which each type of end-to-end error takes place varies as a function of different parameters:

\begin{itemize}
    \item End-to-end errors with different syndromes represent a large percentage of the total number of end-to-end errors when the rate of the code is high. This percentage decreases as the rate of the codes goes from $R_Q = 0.5$ to $R_Q = 0.1$ (see ratio $\frac{\mathrm{E}_1}{\mathrm{E}_T}$ in Figure \ref{fig:results-codes}). This trend becomes further exacerbated as the blocklength of the simulated codes increases, i.e., for low rate large blocklength codes the ratio $\frac{\mathrm{E}_1}{\mathrm{E}_T}$ will be significantly smaller than for low rate short codes. 
    \item End-to-end identical syndrome errors represent the smallest percentage of the total number of end-to-end errors in most of the simulated cases. This is reflected by the fact that $\frac{\mathrm{E}_2}{\mathrm{E}_T} < \frac{\mathrm{E}_1}{\mathrm{E}_T}, \frac{\mathrm{E}_3}{\mathrm{E}_T}$ in all of our simulation outcomes. 
    \item As the noise level of the channel grows, the ratio $\frac{\mathrm{E}_1}{\mathrm{E}_T}$ becomes larger and the ratio $\frac{\mathrm{E}_2}{\mathrm{E}_T}$ diminishes. The ratio of end-to-end degenerate errors $\frac{\mathrm{E}_3}{\mathrm{E}_T}$ stays relatively constant. 

\end{itemize}

The relationships between these parameters and different types of end-to-end errors serve to draw conclusions, some of which should be explored in future work. For instance, the large values of $\frac{\mathrm{E}_1}{\mathrm{E}_T}$ in many of the simulated instances can be understood as a sign that performance gains may be attained by improving the decoding algorithm. This means that applying modified decoding strategies, such as those of \cite{mod-BP, efb, osd1, osd2, refined}, will aid in reducing the presence of end-to-end errors with different syndromes (they estimate the syndrome correctly when the original decoder does not) and improve performance. A matter that should be considered is how often these strategies produce failed error corrections in the form of end-to-end identical syndrome errors. For the methodology of \cite{efb}, such events were shown to be rare, hence we expect these strategies to be a good approach to improve the performance of QLDPC codes. On the other hand, the large values of $\frac{\mathrm{E}_3}{\mathrm{E}_T}$ when compared to $\frac{\mathrm{E}_2}{\mathrm{E}_T}$, especially at higher blocklengths, show that end-to-end identical syndrome errors are the least frequent of all the end-to-end error types. Despite the relatively small percentage that end-to-end identical syndrome errors represent, it is possible that their relevance will grow when the amount of end-to-end errors with different syndromes is reduced (using modified decoding strategies) or when the degenerate content of the code is increased (through design). At this point, it may be that further improvements in performance will only be possible by designing an optimal degenerate decoder with the capability to correct end-to-end identical syndrome errors. Finally, given that the method proposed in this work is valid to detect end-to-end degenerate errors, it may be interesting in future work to employ this methodology to specifically design codes to be degenerate. This could result in code constructions in which the likelihood of end-to-end degenerate errors is maximized, which would allow the positive effects of degeneracy (improved error correction capabilities without a decoding complexity increase) to be completely exploited for quantum error correction purposes.

\section{Conclusion}\label{sec:conclusion}

We have presented a method to detect degenerate errors in sparse quantum codes in a computationally efficient manner. We have also shown how this method is less complex than other existing strategies to compute the logical error rate of sparse CSS quantum codes. Making use of our scheme, we have shown how sparse quantum codes have a significant percentage of degenerate errors. This means that the discrepancy between the logical error rate and the physical error rate is exacerbated for sparse quantum codes. Our results show that, for specific families of QLDPC codes, performance may be up to $20\%$ better than would be expected from previous results in the literature that are based on the physical error rate. In addition, these simulation outcomes serve to show how performance may be improved by constructing degenerate quantum codes, and they also speak toward the positive impact that modified decoding strategies can have on the performance of sparse quantum codes.

\section{Acknowledgements}
This work was financially supported by the Spanish Ministry of Economy and Competitiveness through the ADELE project (Grant  No. PID2019-104958RB-C44), by the Spanish Ministry of Science and Innovation through the proyect ``Few-qubit quantum hardware, algorithms and codes, on photonic and solid- state systems'' (PLEC2021-008251), by the Ministry of Economic Affairs and Digital Transformation of the Spanish Government through the QUANTUM ENIA project call - QUANTUM SPAIN project, by the European Union through the Recovery, Transformation and Resilience Plan - NextGenerationEU within the framework of the Digital Spain 2025 Agenda, and by the Diputación Foral de Gipuzkoa through the DECALOQC Project No. E 190 / 2021 (ES). This work was also funded in part by NSF Award No. CCF-2007689. The authors would like to thank Dr. Pavel Panteleev, Dr. Kao-Yueh Kuo, and Mr. Alex Rigby for their valuable input and useful discussions.


\begin{thebibliography}{00}


\bibitem{degen1} 
D. P. DiVincenzo, P. W. Shor, and J. A. Smolin, ``Quantum-channel capacity of very noisy channels'', \emph{Phys. Rev. A},  vol.57, pp. 830-839, 1998. doi: 10.1103/PhysRevA.57.830.

\bibitem{degen2} 
G. Smith and J. A. Smolin, ``Degenerate Quantum Codes for Pauli Channels'', \emph{Phys.
Rev. Lett.},  vol.98, pp. 030501, 2007. doi: 10.1103/PhysRevLett.98.030501.


\bibitem{degen3} 
D. Poulin and Y. Chung, ``On the iterative decoding of sparse quantum codes'', \emph{QIC}, vol.8, no.10 pp. 987, 2008. doi: 10.5555/2016985.2016993.

\bibitem{degen4} 
M. H. Hsieh and F. Le Gall, ``NP-hardness of decoding quantum error-correction codes'', \emph{Phys. Rev. A}, vol. 83, pp. 052331, 2010. doi: 10.1103/PhysRevA.83.052331.

\bibitem{symplec1} A. R. Calderbank and P. W. Shor, ``Good quantum error correcting codes exist,'' \emph{Phys. Rev. A}, vol. 54, pp. 1098-1105, 1996. doi: 10.1103/PhysRevA.54.1098.

\bibitem{symplec2} A. R. Calderbank, E. M. Rains, P. W. Shor, and N. J. A. Sloane, ``Quantum error correction via codes over GF(4),'' \emph{IEEE Trans. Inf. Theory}, vol. 44, pp.1369–1387, 1998. doi: 10.1109/18.681315.

\bibitem{CSS2} 
A. Steane, Proceedings of The Royal Society A Mathematical, ``Multiple-particle interference and quantum error correction'', \emph{Physical and Engineering Sciences}, vol. 452, no. 1954, 1996. doi: 10.1098/rspa.1996.0136.	

\bibitem{QSC} D. Gottesman, ``Stabilizer codes and quantum error correction,'' \textit{Ph.D. dissertation}, California Inst. Tech., Pasadena, CA, USA, 1997. online: arXiv:quant-ph/9705052.

\bibitem{logical}
B. Yoshida and I. L. Chuang, {``Framework for classifying logical operators in stabilizer codes'',} \emph{Phys. Rev. A}, vol. 81, pp. 052302, 2010. doi: 10.1103/PhysRevA.81.052302.

\bibitem{softVit}
E. Pelchat and D. Poulin, {``Degenerate Viterbi Decoding'',} \emph{IEEE Trans. on Information Theory}, vol. 59, no. 6, pp. 3915 - 3921, 2013. doi: 10.1109/TIT.2013.2246815.

\bibitem{neural}
S. Varsamopoulos, B. Criger, and K. Bertels, {``Decoding small surface codes with feedforward neural networks'',} \emph{Quantum Science and Technology}, vol. 3, no. 1, pp. 015004, 2017. doi: 10.1088/2058-9565/aa955a.

\bibitem{Hard} 
P. Iyer and D. Poulin, ``Hardness of decoding quantum stabilizer codes'', \emph{IEEE Trans. on Information Theory}, vol. 61, no. 9, pp. 5209-5223, 2015. doi: 10.1109/TIT.2015.2422294.

\bibitem{softPoul}
D. Poulin, {``Optimal and Efficient Decoding of Concatenated Quantum Block Codes'',} \emph{Phys. Rev. A}, vol. 74, pp. 052333, 2006. doi: 10.1103/PhysRevA.74.052333.

\bibitem{reviewPat}
P. Fuentes, J. Etxezarreta Martinez, P. M. Crespo, and J. Garcia-Frias, {``Degeneracy and its impact on the decoding of sparse quantum codes'',} \emph{IEEE Access} vol. 9, pp. 89093-89119, 2021. doi: 10.1109/ACCESS.2021.3089829.

\bibitem{osd1}  P. Panteleev and G. Kalachev, “Degenerate quantum LDPC codes with good ﬁnite length performance,” online: arXiv:1904.02703, 2019. 

\bibitem{osd2}  J. Roffe, D. R. White, S. Burton, and E. T. Campbell, “Decoding Across the Quantum LDPC Code Landscape,” \textit{Phys. Rev. Research}, vol. 2, no. 4, pp. 043423, 2020. doi: 10.1103/PhysRevResearch.2.043423.

\bibitem{refined}  K. -Y. Kuo and C. -Y. Lai, “Refined Belief Propagation Decoding of Sparse-Graph Quantum Codes,” \textit{IEEE Journal on Selected Areas in Information Theory}, vol. 1, no. 2, pp. 487-498, 2020. doi: 10.1109/JSAIT.2020.3011758.

\bibitem{exploiting} K.-Y. Kuo and C.-Y. Lai, ``Exploiting degeneracy in belief propagation
decoding of quantum codes'', online: arXiv:2104.13659.

\bibitem{Til} T. Camara, H. Ollivier and J. Tillich, ``A class of quantum LDPC codes: construction and performances under iterative decoding," \textit{2007 IEEE International Symposium on Information Theory}, 2007, pp. 811-815, doi: 10.1109/ISIT.2007.4557324.

\bibitem{mod-BP} 
A. Rigby, J. C. Olivier, and P. Jarvis, ``Modified belief propagation decoders for quantum low-density parity-check codes,” \textit{Physical Review A}, vol. 100, no. 1, pp. 012330, 2019. doi: 10.1103/PhysRevA.100.012330.

\bibitem{jgf1} H. Lou, and J. Garcia-Frias ``Quantum error-correction using codes with low-density generator matrix,'' \emph{IEEE 6th Workshop on Signal Processing Advances in Wireless Communications}, 2005. doi: 10.1109/SPAWC.2005.1506298.


\bibitem{jgf2} 
H. Lou and J. Garcia-Frias, ``On the Application of Error-Correcting Codes with Low Density Generator Matrix over Different Quantum Channels'', \textit{4th International Symposium on Turbo Codes \& Related Topics}, 2006.
online: https://ieeexplore.ieee.org/document/5755950.

\bibitem{qldpc15} Z. Babar, P. Botsinis, D. Alanis, S. X. Ng, and L. Hanzo, ``Fifteen Years of Quantum LDPC Coding and Improved Decoding Strategies,'' \textit{IEEE Access}, vol. 3, pp. 2492-2519, 2015. doi. 10.1109/ACCESS.2015.2503267.


\bibitem{efb}
Y. Wang, B. C. Sanders, B. Bai and X. Wang, ``Enhanced Feedback Iterative Decoding of Sparse Quantum Codes,” \textit{IEEE Trans. Inform. Theory}, vol. 58, no. 2, pp. 1231-1241, 2012. doi: 10.1109/TIT.2011.2169534.

\bibitem{bab1} Z. Babar, P. Botsinis, D. Alanis, S. Xin Ng and L. Hanzo, ``Construction of Quantum LDPC Codes From Classical Row-Circulant QC-LDPCs," \textit{IEEE Communications Letters}, vol. 20, no. 1, pp. 9-12, 2016, doi: 10.1109/LCOMM.2015.2494020.


\bibitem{patrick1} P. Fuentes, J. Etxezarreta Martinez, P. M. Crespo, and J. Garcia-Frias, ``Approach for the construction of non-Calderbank-Steane-Shor low-density-generator-matrix based quantum codes,'' \textit{Phys. Rev. A}, vol. 102, pp. 012423, 2020. doi:10.1103/PhysRevA.102.012423.

\bibitem{QEClidar} D. Lidar, and T. Brun, ``Quantum Error Correction,'' \textit{Cambridge: Cambridge University Press}, 2013. doi: 10.1017/CBO9781139034807.

\bibitem{trapping-sets} N. Raveendran and B. Vasić, “Trapping Sets of Quantum LDPC Codes,” \textit{Quantum}, vol. 5, no. 562, 2021. doi: 10.22331/q-2021-10-14-562.

\bibitem{NielsenChuang} M. A. Nielsen, and I. Chuang, ``Quantum Computation and Quantum Information: 10th Anniversary Edition,'' \emph{Cambridge: Cambridge University Press}, 2011. doi:10.1017/CBO9780511976667.



\bibitem{EAQECC}
T. Brun, I. Devetak, and M. Hsieh, {``Correcting Quantum Errors with Entanglement'',} \textit{Science}, vol. 314, no. 5798, pp. 436-439, 2006. doi: 	10.1126/science.1131563.

\bibitem{josurev} J. Etxezarreta Martinez, P. Fuentes, P. M. Crespo, and J. Garcia-Frias, ``Approximating Decoherence Processes for the Design and Simulation of Quantum Error Correction Codes in Classical Computers,'' \emph{IEEE Access}, vol. 8, pp. 172623-172643, 2020. doi: 10.1109/ACCESS.2020.3025619.

\bibitem{catalytic}
 T. A. Brun, I. Devetak, and M. Hsieh, ``Catalytic Quantum Error Correction,'' \emph{IEEE Trans. Inf. Theory}, vol. 60, no. 6, pp. 3073-3089, 2014. doi: 10.1109/TIT.2014.2313559.
 
 
\bibitem{classicaltoquantum} Z. Babar, D. Chandra, H. V. Nguyen, P. Botsinis, D. Alanis, S. X. Ng, and L. Hanzo, ``Duality of Quantum and Classical Error Correction Codes: Design Principles and Examples,'' \emph{IEEE Communications Surveys Tutorials}, vol. 29, no. 1, pp. 970--1010, 2019.  doi:10.1109/COMST.2018.2861361.

\bibitem{Pat-thesis} P. Fuentes, ``Error Correction for Reliable Quantum Computing,'' \textit{Ph.D. dissertation}, Tecnun-School of Engineering, Donostia - San Sebastian, Spain, 2022. online: arXiv:2202.08599.


\bibitem{EAQTC} M. M. Wilde, M. Hsieh, and  Z. Babar, ``Entanglement-Assisted Quantum Turbo Codes,'' \emph{IEEE Trans. Inf. Theory}, vol. 60, no. 2, pp.  1203--1222, 2014. doi: 10.1109/TIT.2013.2292052.

\bibitem{toricphd1} D. K. Tuckett, A. S. Darmawan, T. C. Chubb, S. Bravyi, S. D. Bartlett, and S. T. Flammia, "Tailoring Surface Codes for Highly Biased Noise," \textit{Phys. Rev. X}, vol. 9, no. 4, pp. 041031, 2019. doi: 10.1103/PhysRevX.9.041031.

\bibitem{toricphd2} D. K. Tuckett, {``Tailoring surface codes: Improvements in quantum error correction with biased noise'',} \textit{PhD. thesis}, University of Sydney, 2020.

\bibitem{sabo} E. Sabo, A. B. Aloshious, and K. R. Brown, “Trellis Decoding For Qudit Stabilizer Codes And Its Application To Qubit Topological Codes,” online: arXiv:2106.08251, 2021.

\bibitem{BP} 
J. Pearl, \textit{Probabilistic Reasoning in Intelligent Systems: Networks of Plausible Inference}, Morgan Kauffman, 1988. online: https://www.sciencedirect.com/book/.

\bibitem{spa} 
F.R. Kschischang, B.J. Frey, and H.A. Loeliger, ``Factor graphs and the sum-product algorithm'', \emph{IEEE Transactions on Information Theory},  vol.47, no. 2, pp. 498--519, 2001. doi: 10.1109/18.910572.


\bibitem{Benny} K. Kuo and C. Lu, "A further study on the encoding complexity of quantum stabilizer codes," \textit{2010 International Symposium On Information Theory \& Its Applications}, 2010, pp. 1041-1044, doi: 10.1109/ISITA.2010.5649496.

\bibitem{Wilde} M. Wilde, "Logical operators of quantum codes," \textit{Phys. Rev. A}, vol. 79, pp. 062322, 2009. doi: 10.1103/PhysRevA.79.062322.

\bibitem{Mac} D. J. C. MacKay, “Good error-correcting codes based on very sparse matrices,” \textit{IEEE Trans. Inf. Theory}, vol. 45, pp. 399–431, 1999. doi: 10.1109/18.748992.


\bibitem{CSS1} 
A. R. Calderbank and P. W. Shor, ``Good quantum error correcting codes exist,'' \textit{Physical Review A}, vol. 54, pp. 1098-1105, 1996. doi: 10.1103/PhysRevA.54.1098.

\bibitem{complex}
T. J. Richardson and R. L. Urbanke, ``Efficient encoding of low-density parity-check codes,'' \textit{IEEE Transactions on Information Theory}, vol. 47, no. 2, pp. 638-656, 2001. doi: 10.1109/18.910579.

\bibitem{G1}
S. Myung, K. Yang and J. Kim, ``Quasi-cyclic LDPC codes for fast encoding,'' \textit{IEEE Transactions on Information Theory}, vol. 51, no. 8, pp. 2894-2901, 2005, doi: 10.1109/TIT.2005.851753.

\bibitem{G2}
M. Battaglioni, P. Santini, M. Baldi and G. Cancellieri, "Obtaining structured generator matrices for QC-LDPC codes," \textit{2019 AEIT International Annual Conference (AEIT)}, pp. 1-6, 2019. doi:10.23919/AEIT.2019.8893395.

\bibitem{G3}
M. Baldi, M. Bianchi, G. Cancellieri, F. Chiaraluce and T. Klove, ``On the generator matrix of array LDPC codes,'' \textit{SoftCOM 2012 20th International Conference on Software, Telecommunications and Computer Networks}, pp. 1-5, 2012.

\bibitem{G4}
Su-Chang Chae and Yun-Ok Park, ``Low complexity encoding of improved regular LDPC codes,'' \textit{IEEE 60th Vehicular Technology Conference}, vol. 4, pp. 2535-2539, 2004. doi: 10.1109/VETECF.2004.1400513.

\bibitem{IRA}
H. Jin, A. D. Kh, R. J. McEliece, ``Irregular repeat-accumulate codes'', \textit{Proc. of 2nd International Symposium on Turbo Codes and Related Topics}, Brest, France, 2000. 

\bibitem{patrick2} P. Fuentes, J. Etxezarreta Martinez, P. M. Crespo, and J. Garcia-Frias, ``Design of LDGM-based quantum codes for asymmetric quantum channels,'' \textit{Phys. Rev. A}, vol. 103, pp. 022617, 2021. doi: 10.1103/PhysRevA.103.022617.

\bibitem{patrick3}
P. Fuentes, J. Etxezarreta Martinez, P. M. Crespo, and J. Garcia-Fr\'ias, ``Performance of non-CSS LDGM-based quantum codes over the Misidentified Depolarizing Channel,'' \textit{IEEE International Conference on Quantum Computing and Engineering (QCE20)}, 2020. doi:10.1109/QCE49297.2020.00022.

\bibitem{MonteCarlo} M. Jeruchim, ``Techniques for Estimating the Bit Error Rate in the Simulation of Digital Communication Systems,'' \emph{IEEE J. Selected Areas Commun,} {\bf 1984}, {\em 2},~153--170.



\end{thebibliography}
\end{document}